# ON THE ONE METHOD OF A THIRD-DEGREE BEZIER TYPE SPLINE CURVE CONSTRUCTION

Stelia O., Potapenko L., Sirenko I.

Faculty of Computer Sciences and Cybernetics,
Taras Shevchenko National University of Kyiv, Kyiv, Ukraine
oleg.stelya@gmail.com, lpotapenko@ukr.net,
i.sirenko@gmail.com

**Abstract.** A method is proposed for constructing a spline curve of the Bezier type, which is continuous along with its first derivative by a piecewise polynomial function. Conditions for its existence and uniqueness are given. The constructed curve lies inside the convex hull of the control points, and the segments of the broken line connecting the control points are tangent to the curve. To construct the curve, we use the approach proposed earlier for constructing a parabolic spline. The idea is to use additional points with unknown values of some function. Additional points are used as spline nodes, and the function values are determined from the condition of the first derivative continuity of a piecewise polynomial curve. In multiple interpolation nodes, the function takes the given values and the values of the first derivative, which are determined by the control points. Examples of constructing a spline curve are given.

## 1   Introduction

The history of constructing and using smooth curves is quite old. The algorithms for constructing piecewise polynomial smooth curves of low degrees received rapid development with the advent and development of computer technology. On the one hand, computers can significantly accelerate the construction of such curves for various technical applications, and on the other hand the computers themselves are "consumers" of various algorithms in computer graphics systems [1, 2]. The most frequently Bezier curves, spline curves and B-splines are used for these purposes.

Curves subsequently called Bezier curves [3, 4, 5, 6] were developed in the 1960s independently by Pierre Bezier of the Renault car company and Paul de Castillo of Citroën, where they used for the design of car bodies. Bezier curves are used in computer graphics to draw smooth bends, animations and other applications.

The wide application of Bezier curves for approximation problems is related with their convenience as an analytical description and a visual geometric construction. With respect to the computer graphics, this means that the user can set the shape of the curve interactively, i. e. moving the curve-defining points on the screen. Also Bezier curves can be used in cartography when it is required to build a route on the map with given initial and end points and which passes near some points

In work [7], an algorithm is given for calculating the coefficients of a parametric spline, which approximates the sequence of experimental data. The Bezier curves were used as the approximating spline.

B-splines [4, 5, 8] is a generalization of Bezier curves. Modeling on the basis of inhomogeneous rational B-splines has the advantage over other methods. With the help of such splines it is easier to simulate the surfaces of natural objects, as well as surfaces that have complexly curved profiles. The models of such splines provide the best quality of the rounded edges of objects visualization [8].

In the book [9] an algorithmic processing of Bézier curves and spline curves is considered.

In this publication, we propose and justify an algorithm for constructing a curve that has a number of properties of both Bezier curves and spline curves. In connection with this, the resulting curve can be called a third-order spline curve of Bezier type.

## 2    A constructive method for proving the existence and uniqueness of a spline curve

We consider a certain interval $[a,b]$, on which we define the grid

$$\Delta_\tau : a = \tau_1 < \tau_2 < ... < \tau_N = b.$$

At the nodes $\tau_i$ we set the values $F_i$. To construct the curve, we use the approach proposed in [10] for parabolic splines. Along with the grid $\Delta_\tau$, we introduce a grid $\Delta_x : \tau_1 = x_1 < x_2 < ... < x_{N+1} = \tau_N$, where $\tau_{i-1} < x_i < \tau_i$. The values of a certain function at the points $x_i$ will be denoted by $f_i$.

We introduce the notation $h_i = \tau_i - \tau_{i-1}$, $\mu_i = \tau_i - x_i$

Then $x_i - \tau_{i-1} = h_i - \mu_i$. Taking these notations into account, we get:
$f_i = [F_{i-1}(h_i - \mu_i) + F_i \mu_i]/h_i$, $f_i' = (F_i - F_{i-1})/h_i$.

We will construct a spline curve of the third order, for which the points $\tau_i$ will be nodes of the spline, and the points $x_i$ will be multiples of the interpolation nodes. To construct the curve, we construct a cubic spline of defect 2 on the interval $[a,b]$, satisfying the following conditions:

$$S^3(x_i) = f_i, \tag{1}$$

$$(S^3)'(x_i) = f_i', \quad i = \overline{2,N}. \tag{2}$$

We denote $\varphi_i$, $i = \overline{1,N}$ the unknown values of the function $S^3(x)$ at the nodes $\tau_i$ of the spline,

**Theorem.** Under the conditions $\frac{1}{3} \leq \alpha_i \leq \frac{2}{3}$, where $\alpha_i = \frac{\mu_i}{h_i}$, the interpolation cubic spline of defect 2 $S^3(x)$ for grids $\Delta_x$, $\Delta_\tau$ on the interval $[a,b]$, satisfying conditions (1) - (2), there exists and unique one.

Proof. To construct a spline, we write the third-order interpolation Hermite polynomial [11] on each of the intervals $[\tau_{i-1}, \tau_i]$, $i = \overline{2, N-1}$.

$$S^3(x) = \varphi_{i-1} \frac{(x-x_i)(x-\tau_i)}{(\tau_{i-1}-x_i)(\tau_{i-1}-\tau_i)} + \varphi_i \frac{(x-x_i)(x-\tau_{i-1})}{(\tau_i-x_i)(\tau_i-\tau_{i-1})} + $$
$$+ f_i \frac{(x-\tau_i)(x-\tau_{i-1})}{(x_i-\tau_i)(x_i-\tau_{i-1})} + Q_1(x-\tau_i)(x-\tau_{i-1})(x-x_i) \quad (3)$$

for $x \in [\tau_{i-1}, \tau_i]$,

$$S^3(x) = \varphi_i \frac{(x-x_{i+1})(x-\tau_{i+1})}{(\tau_i-x_{i+1})(\tau_i-\tau_{i+1})} + \varphi_{i+1} \frac{(x-x_{i+1})(x-\tau_i)}{(\tau_{i+1}-x_{i+1})(\tau_{i+1}-\tau_i)} + $$
$$+ f_{i+1} \frac{(x-\tau_i)(x-\tau_{i+1})}{(x_{i+1}-\tau_i)(x_{i+1}-\tau_{i+1})} + Q_2(x-\tau_i)(x-\tau_{i+1})(x-x_{i+1}) \quad (4)$$

for $x \in [\tau_i, \tau_{i+1}]$.

It is obvious from the relations (3) - (4) that the condition (1) is satisfied.

To determine the quantities $Q_1$ and $Q_2$ we use condition (2), i.e. the following relations must be satisfied:

$$(S^3)'(x_i) = f_i', \quad (S^3)'(x_{i+1}) = f_{i+1}'.$$

We consider $(S^3)'(x)$ for $x \in [\tau_{i-1}, \tau_i]$

$$(S^3)'(x) = \varphi_{i-1} \frac{(x-x_i)+(x-\tau_i)}{(\tau_{i-1}-x_i)(\tau_{i-1}-\tau_i)} + \varphi_i \frac{(x-x_i)+(x-\tau_{i-1})}{(\tau_i-x_i)(\tau_i-\tau_{i-1})} + $$
$$+ f_i \frac{(x-\tau_i)+(x-\tau_{i-1})}{(x_i-\tau_i)(x_i-\tau_{i-1})} + Q_1\{(x-x_i)[(x-\tau_i)+(x-\tau_{i-1})]+(x-\tau_i)(x-\tau_{i-1})\}. \quad (5)$$

Similarly we consider $(S^3)'(x)$ for $x \in [\tau_i, \tau_{i+1}]$

$$(S^3)'(x) = \varphi_i \frac{(x-x_{i+1})+(x-\tau_{i+1})}{(\tau_i-x_{i+1})(\tau_i-\tau_{i+1})} + \varphi_{i+1} \frac{(x-x_{i+1})+(x-\tau_i)}{(\tau_{i+1}-x_{i+1})(\tau_{i+1}-\tau_i)} + $$
$$+ f_{i+1} \frac{(x-\tau_i)+(x-\tau_{i+1})}{(x_{i+1}-\tau_i)(x_{i+1}-\tau_{i+1})} + Q_2\{(x-x_{i+1})[(x-\tau_i)+(x-\tau_{i+1})]+(x-\tau_i)(x-\tau_{i+1})\}. \quad (6)$$

$$(S^3)'(x_i) = \varphi_{i-1} \frac{x_i - \tau_i}{(\tau_{i-1} - x_i)(\tau_{i-1} - \tau_i)} + \varphi_i \frac{x_i - \tau_{i-1}}{(\tau_i - x_i)(\tau_i - \tau_{i-1})} +$$

$$f_i \frac{(x_i - \tau_i) + (x_i - \tau_{i-1})}{(x_i - \tau_i)(x_i - \tau_{i-1})} + Q_1(x - \tau_i)(x - \tau_{i-1}) = \frac{F_i - F_{i-1}}{h_i},$$

$$(S^3)'(x_{i+1}) = \varphi_i \frac{x_{i+1} - \tau_{i+1}}{(\tau_i - x_{i+1})(\tau_i - \tau_{i+1})} + \varphi_{i+1} \frac{x_{i+1} - \tau_i}{(\tau_{i+1} - x_{i+1})(\tau_{i+1} - \tau_i)} +$$

$$+ f_{i+1} \frac{(x_{i+1} - \tau_i) + (x_{i+1} - \tau_{i+1})}{(x_{i+1} - \tau_i)(x_{i+1} - \tau_{i+1})} + Q_2(x_{i+1} - \tau_i)(x_{i+1} - \tau_{i+1}) = \frac{F_{i+1} - F_i}{h_{i+1}}.$$
(7)

Using the above notation for the steps $h_i$ and $\mu_i$, we obtain:

$$-\varphi_{i-1} \frac{\mu_i}{(h_i - \mu_i)h_i} + \varphi_i \frac{h_i - \mu_i}{h_i \mu_i} - f_i \frac{h_i - 2\mu_i}{(h_i - \mu_i)\mu_i} - Q_1(h_i - \mu_i)\mu_i = \frac{F_i - F_{i-1}}{h_i},$$
(8)

wherefrom

$$Q_1 = -\varphi_{i-1} \frac{1}{(h_i - \mu_i)^2 h_i} + \varphi_i \frac{1}{h_i \mu_i^2} - f_i \frac{h_i - 2\mu_i}{(h_i - \mu_i)^2 \mu_i^2} - \frac{F_i - F_{i-1}}{h_i \mu_i (h_i - \mu_i)}.$$
(9)

Similarly

$$-\varphi_i \frac{\mu_{i+1}}{(h_{i+1} - \mu_{i+1})h_{i+1}} + \varphi_{I+1} \frac{h_{i+1} - \mu_{i+1}}{h_{i+1} \mu_{i+1}} - f_{i+1} \frac{h_{i+1} - 2\mu_{i+1}}{(h_{i+1} - \mu_{i+1})\mu_{i+1}} -$$

$$- Q_2(h_{i+1} - \mu_{i+1})\mu_{i+1} = \frac{F_{i+1} - F_i}{h_{i+1}},$$
(10)

wherefrom we get

$$Q_2 = -\varphi_i \frac{1}{(h_{i+1} - \mu_{i+1})^2 h_{i+1}} + \varphi_{i+1} \frac{1}{h_{i+1} \mu_{i+1}^2} -$$

$$- f_{i+1} \frac{h_{i+1} - 2\mu_{i+1}}{(h_{i+1} - \mu_{i+1})^2 \mu_{i+1}^2} - \frac{F_{i+1} - F_i}{h_{i+1} \mu_{i+1} (h_{i+1} - \mu_{i+1})}.$$
(11)

To ensure the smoothness of the resulting curve, that is, the continuity of the first derivative, we require the fulfillment of the relation $(S^3)'(\tau_i - 0) = (S^3)'(\tau_i + 0)$, where

$$(S^3)'(\tau_i - 0) = \varphi_{i-1} \frac{\mu_i}{(h_i - \mu_i)h_i} + \varphi_i \frac{h_i + \mu_i}{h_i \mu_i} - f_i \frac{h_i}{(h_i - \mu_i)\mu_i} + Q_1 \mu_i h_i,$$
(12)

$$(S^3)'(\tau_i + 0) = -\varphi_i \frac{2h_{i+1} - \mu_{i+1}}{(h_{i+1} - \mu_{i+1})h_{i+1}} - \varphi_{I+1} \frac{h_{i+1} - \mu_{i+1}}{h_{i+1}\mu_{i+1}} +$$
$$+ f_{+1I} \frac{h_{i+1}}{(h_{i+1} - \mu_{i+1})\mu_{i+1}} - Q_2(h_{i+1} - \mu_{i+1})h_{i+1}. \tag{13}$$

Equating expressions (12) and (13) and substituting the values $Q_1$ and $Q_2$ from (9) and (11), we obtain a system of linear algebraic equations with a tridiagonal matrix for determining $\varphi_i$:

$$\varphi_{i-1} \frac{\mu_i}{(h_i - \mu_i)h_i} + \varphi_i \frac{h_i + \mu_i}{h_i\mu_i} - f_i \frac{h_i}{(h_i - \mu_i)\mu_i} - \varphi_{i-1} \frac{\mu_i}{(h_i - \mu_i)^2} + \frac{\varphi_i}{\mu_i} -$$
$$- f_i \frac{(h_i - 2\mu_i)h_i}{(h_i - \mu_i)^2 \mu_i} - \frac{F_i - F_{i-1}}{(h_i - \mu_i)} = -\varphi_i \frac{2h_{i+1} - \mu_{i+1}}{(h_{i+1} - \mu_{i+1})h_{i+1}} - \varphi_{i+1} \frac{h_{i+1} - \mu_{i+1}}{h_{i+1}\mu_{i+1}} +$$
$$+ f_{i+1} \frac{h_{i+1}}{(h_{i+1} - \mu_{i+1})\mu_{i+1}} - \frac{\varphi_i}{(h_{i+1} - \mu_{i+1})} + \varphi_{i+1} \frac{(h_{i+1} - \mu_{i+1})}{\mu_{i+1}^2} - \tag{14}$$
$$- f_{i+1} \frac{(h_{i+1} - 2\mu_{i+1})h_{i+1}}{(h_{i+1} - \mu_{i+1})\mu_{i+1}^2} - \frac{F_{i+1} - F_i}{\mu_{i+1}}.$$

We write expression (14) in the following form:

$$A\varphi_{i-1_i} - (B_1 + B_2)\varphi_{i_i} + C\varphi_{i+1_i} = \Phi_i, \tag{15}$$

where

$$A = \frac{\mu_i^2}{(h_i - \mu_i)^2 h_i}, \quad B_1 = \frac{2h_i + \mu_i}{\mu_i h_i}, \quad B_2 = \frac{3h_{i+1} - \mu_{i+1}}{(h_{i+1} - \mu_{i+1})h_{i+1}}, \quad C = \frac{(h_{i+1} - \mu_{i+1})^2}{h_{i+1}\mu_{i+1}^2}, \tag{16}$$

$$\Phi_i = -f_i \frac{h_i}{(h_i - \mu_i)\mu_i} - f_i \frac{(h_i - 2\mu_i)h_i}{(h_i - \mu_i)^2 \mu_i} - \frac{F_i - F_{i-1}}{(h_i - \mu_i)} + f_{i+1} \frac{h_{i+1}}{(h_{i+1} - \mu_{i+1})\mu_{i+1}} -$$
$$- f_{i+1} \frac{(h_{i+1} - 2\mu_{i+1})h_{i+1}}{(h_{i+1} - \mu_{i+1})\mu_{i+1}^2} - \frac{F_{i+1} - F_i}{\mu_{i+1}}. \tag{17}$$

To close up the system of equations, we add the conditions:

$$\varphi_1 = f_1, \quad \varphi_N = f_N. \tag{18}$$

Obviously, the values $A, B_1, B_2, C$ are positive and $0 < \alpha_i < 1$. If $B_1 > A$ and $B_2 > C$, then system (15) will have a diagonal predominance.

The fulfillment of the condition $B_1 > A$ leads to an inequality $\frac{2h_i + \mu_i}{\mu_i} > \frac{q_i^2}{(h_i - \mu_i)^2}$, from which follows the inequality $\frac{2 + \alpha_i}{\alpha_i} > (\frac{\alpha_i}{1 - \alpha_i})^2$, the solution of which is $\alpha_i < \frac{2}{3}$

Similarly, the fulfillment of the condition $B_2 > C$ leads to an inequality $\frac{3h_{i+1} - \mu_{i+1}}{h_{i+1} - \mu_{i+1}} > \frac{(h_{i+1} - \mu_{i+1})^2}{\mu_{i+1}^2}$, from which follows the inequality $\frac{3 - \alpha_{i+1}}{1 - \alpha_{i+1}} > (\frac{1 - \alpha_{i+1}}{\alpha_{i+1}})^2$, the solution of which is $\alpha_{i+1} > \frac{1}{3}$.

If $\alpha_i = \frac{1}{3}$, then $B_1 > A$ and $B_2 = C$. If $\alpha_i = \frac{2}{3}$, then $B_1 = A$ and $B_2 > C$, wherefrom $B_1 + B_2 > A + C$

Thus, the system of equations (15) - (18) has a diagonal predominance, from which implies the existence and uniqueness of its solution [12]. The theorem is proved.

## 3  Examples of calculations

Let us illustrate the computational properties of the obtained curve using the following examples.

**Example 1**. Let a net function be defined on the interval $[1, 11]$:

| $x_i$ | 1 | 2 | 3 | 4 | 5 | 6 | 7   | 8 | 9  | 10 | 11  |
|-------|---|---|---|---|---|---|-----|---|----|----|-----|
| $F_i$ | 1 | 3 | 3 | 1 | 2 | 7 | 1.5 | 1 | 10 | 2  | 1.5 |

The points $(x_i, F_i)$, by analogy with the terminology accepted for Bezier curves, will be called control. Figure 1 shows the results of calculating the spline curve according to the proposed algorithm. In the figure, the control points are indicated by dots, the broken line connecting them are indicated by dashed line. The solid line is the spline curve constructed. It can be seen from the graph that the constructed curve lies inside the convex hull formed by the control points.

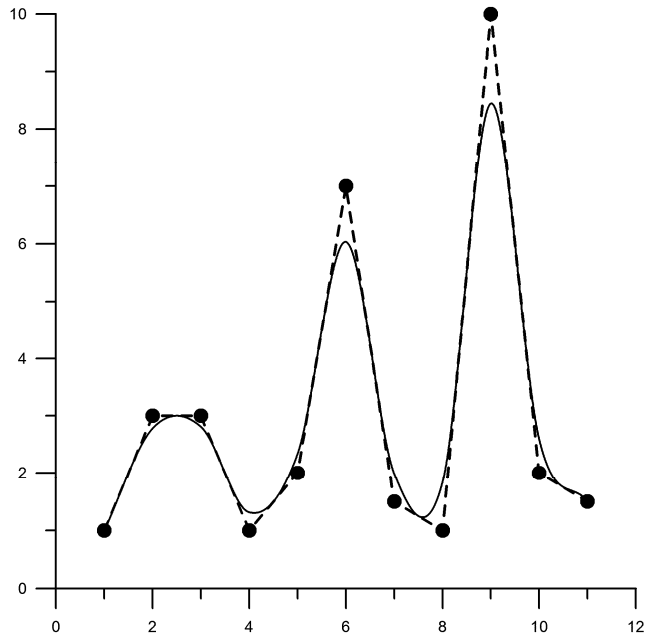

**Fig. 1.**

**Example 2.** Consider a grid function on a segment $[0, 1]$:

| $x_i$ | 0 | 0.1 | 0.2 | 0.3 | 0.4 | 0.5 | 0.6 | 0.7 | 0.8 | 0.9 | 1 |
|---|---|---|---|---|---|---|---|---|---|---|---|
| $F_i$ | 0 | 0.3 | 0.4 | 0.458257569 | 0.489897949 | 0.5 | 0.4898979495 | 0.458257569 | 0.4 | 0.3 | 0 |

The control points lie on the semicircle given by the equation: $y = \sqrt{x - x^2}$, $0 \leq x \leq 1$. The results of the construction are shown in Figure 2. The notation in this figure corresponds to the notation in Figures 1. This example demonstrates a good possibility of approximating of the semicircle arc using the proposed curve.

## 4   Conclusions

An algorithm for constructing and justifying a third degree spline curve of Bezier type is given. The resulting curve has the properties of both a spline and a Bezier curve. The algorithm allows us to construct a uniform piecewise polynomial function that is continuous along with its first derivative on the whole interval. The curve has the third degree, regardless of the number of control points. The above examples demonstrate good approximation properties of the curve, which, according to the authors, can find application in computer graphics systems.

8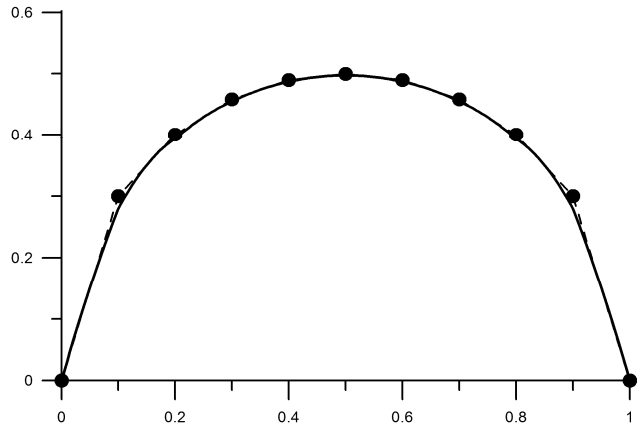

**Fig. 2.**